\documentclass[12pt]{article}
\usepackage[left=2.50cm, right=2.50cm, top=2.50cm, bottom=2.50cm]{geometry}
\usepackage[utf8]{inputenc}
\usepackage[english]{babel}
\usepackage{physics}
\usepackage{hyperref}
\usepackage{amsthm}
\usepackage{mathtools}
\usepackage{graphicx} 
\usepackage{changepage}
\usepackage{verbatim}
\usepackage{empheq}
\usepackage{xcolor}
\numberwithin{equation}{section}
\hypersetup{hidelinks}

\newcommand{\ka}{\kappa}
\newcommand{\kat}{\tilde{\kappa}}
\newcommand{\D}{\partial}
\newcommand{\al}{\alpha}
\newcommand{\alt}{{\tilde{\alpha}}}
\newcommand{\ga}{\gamma}
\newcommand{\gat}{{\tilde\gamma}}
\newcommand{\la}{\lambda}
\newcommand{\lat}{\tilde{\lambda}}
\newcommand{\La}{\Lambda}
\newcommand{\Lat}{\tilde{\Lambda}}
\newcommand{\be}{\beta}
\newcommand{\bet}{{\tilde{\beta}}}
\newcommand{\Par}{\mathcal{P}}
\newcommand{\T}{\mathcal{T}}
\newcommand{\At}{\tilde{A}}

\newcommand{\Gt}{\tilde{G}}
\newcommand{\xit}{\tilde{\xi}}

\newcommand{\Lag}{\mathcal{L}}
\newcommand{\Sa}{S^{^{(1)}}_{3D}}
\newcommand{\Sb}{S^{^{(2)}}_{3D}}

\begin{document}
\noindent

{\bf
{\Large

Holographic Projection of Electromagnetic \\ 

\noindent
Maxwell Theory 
}} 

\vspace{.5cm}
\hrule

\vspace{1cm}

\noindent

{\large\bf{Erica Bertolini\footnote{\tt erica.bertolini@ge.infn.it } and Nicola Maggiore\footnote{\tt nicola.maggiore@ge.infn.it }\\[1cm]}}

\setcounter{footnote}{0}

\noindent
{{}Dipartimento di Fisica, Universit\`a di Genova,\\
via Dodecaneso 33, I-16146, Genova, Italy\\
and\\
{} I.N.F.N. - Sezione di Genova\\
\vspace{1cm}

\noindent
{\tt Abstract~:}
\\The 4D Maxwell theory with single-sided planar boundary is considered. As a consequence of the presence of the boundary, two broken Ward identities are recovered, which, on-shell, give rise to two conserved currents living on the edge.  A Ka\c{c}-Moody algebra formed by a subset of the bulk fields is obtained with central charge proportional to the inverse of the Maxwell coupling constant, and the degrees of freedom of the boundary theory are identified as two vector fields, also suggesting that the 3D theory should be a gauge theory. Finally the holographic contact between bulk and boundary theory is reached in two inequivalent ways, both leading to a unique 3D action describing a new gauge theory of two coupled vector fields with a topological Chern-Simons term with massive coefficient. In order to check that the 3D projection of 4D Maxwell theory is well defined, we computed the energy-momentum tensor and the propagators. The role of discrete symmetries  is briefly discussed. \\[10px]

\newpage

\section{Introduction}

Boundaries exist in Nature. Their presence is usually swept under the rug, when teaching classes, except then saying, rather vaguely indeed, that ``boundary effects'' should be taken into account, which substantially affect  the idealized bulk-only theories. Think, for instance, to the inexistent ``infinitely long'' solenoids, or to the ideal ``infinitely extended'' parallel plates of a capacitor. The Casimir effect \cite{Casimir:1948dh} perhaps is the first highly nontrivial example of boundary effect which has been thoroughly studied in a systematic way. The role of boundaries has been largely discussed in 2D Conformal Field Theory \cite{Moore:1989yh, Cardy:2004hm}. In particular, in \cite{Moore:1989yh} the zoo of Conformal Field Theories has been tamed by means of a boundary put on the 3D topological Chern-Simons (CS) theory. In \cite{Cardy:2004hm}, instead, the role of the boundary, and in particular of the boundary conditions, has been exploited for the study of the Virasoro algebras and their extensions (Ka\c{c}-Moody, superconformal, W-algebras). In field theory, the pioneering work which must be referred to is \cite{Symanzik:1981wd}, where Symanzik  gave the first formulation of field theory with boundary, defined as the surface which separates propagators, $i.e.$ two-points Green functions, in the sense that propagators computed between points lying on opposite sides of the boundary must vanish. This approach relies on very general principles of field theory, like locality and power counting, and not much space is left to arbitrariness. For instance, the conditions which must by fulfilled by the quantum fields on the boundary  (of the Dirichlet, Neumann or Robin type), are not put by hand in the theory, but are those which naturally come out from the request of separability of propagators. This approach has been very fruitful in the study of Topological Field Theories (TFT) with planar boundary \cite{Amoretti:2013nv}. TFT are characterized by the absence of physical local observables, the only observables being global properties of the manifold where they are built, like the genus, or the numbers of holes and handles \cite{Birmingham:1991ty}. In other words, TFT have vanishing Hamiltonian and energy-momentum tensor, and it is rather surprising that for such non-physical theories it has been possible to establish
\cite{Blasi:1990pf,Blasi:1990bk,Emery:1991tf}
that on their lower dimensional edge, conserved currents exist, which form Ka\c{c}-Moody algebras 
\cite{Kac:1967jr,Moody:1966gf}, whose central charge is inversely proportional to the coupling constant of the bulk theory, and directly related to the velocity of the boundary propagating Degrees Of Freedom (DOF). This property seems to be a common feature of different physical situations, like the 3D Fractional Quantum Hall Effect \cite{Cappelli:2018dti, Blasi:2008gt} 
and the Topological Insulators in 3D \cite{Cho:2010rk,Cappelli:2016xwp,Blasi:2011pf} and 4D \cite{Schnyder:2008tya,Fu:2007uya,Amoretti:2012hs}. 
The boundary conditions, which are not imposed, as we said, play a very important role in the identification of the nature of the edge DOF. Experimentally indeed, the edge states of the Fractional Quantum Hall Effect and of the Topological Insulators are fermionic, while the corresponding bulk theories are completely bosonic, being described in terms of gauge fields. But the boundary conditions have been recognized to be the conditions for the fermionization of bosonic DOF \cite{Aratyn:1984jz, Aratyn:1983bg, Amoretti:2013xya}. In the Symanzik's approach, the boundary separates two half-spaces: left and right hand side with respect to a plane. Single-sided boundaries can also be considered, which correspond to quite different physical situations from the cases previously described. Think for instance to the AdS/CFT correspondence \cite{Witten:1998qj,Polchinski:2010hw,Klebanov:2000me}, which is exactly of that type, showing dualities between D-dimensional gravity bulk theories and their holographic counterparts on their  (D-1) boundaries, where the extra ``energy'' dimension run from zero to infinity. The AdS/CFT holographic correspondence, a.k.a. gauge/gravity duality, originally conjectured in string theory \cite{Maldacena:1997re}, later received much attention in condensed matter theory, enough to introduce for that case a new acronym (AdS/CMT). The bulk/boundary correspondence turned out to be a powerful new technique to study strongly coupled systems, reviewed for instance in \cite{Sachdev:2010ch,Hartnoll:2009sz,Zaanen:2015oix,Amoretti:2017xto}.  The gauge/gravity duality falls in the more general topic of field theories with boundary, and one may indeed refers to holography without gravity \cite{2013ass,Amoretti:2014kba}. 
To avoid possible misunderstandings, we stress that the holography we are dealing with in this paper does not concern the strong-weak coupling duality of the AdS/CFT correspondence. The 4D Maxwell theory without matter is, in fact, a free theory without interaction. Therefore we can calculate anything exactly or, in some sense, the model is trivial. There does not exist any non-perturbative effect like strong-weak duality.
In the single-sided case, Symanzik's separation requirement on propagators does not seem to be the most natural one. It is easier, and more intuitive, to implement the confinement of the theory in a half-space by means of a theta Heaviside step function directly introduced in the bulk action \cite{Amoretti:2014iza,Maggiore:2017vjf}. 
Keeping strict the request of not imposing particular boundary conditions, these can be found by means of a kind of variational principle on the equations of motion. The theories with singe-sided boundaries are therefore treated with a different approach, and the physical results on the boundary do not necessarily coincide with those of the separating boundaries. This is the case, for instance, of 3D Maxwell-Chern-Simons theory, where the Maxwell term is completely transparent in case of double-sided boundary \cite{Blasi:2010gw} and algebraically active in the single-sided case \cite{Maggiore:2018bxr,Geiller:2019bti}. It is precisely the different role of the non-topological Maxwell term that motivated the study of non-TFT, or theories with non-topological terms, defined in half-spaces \cite{Blasi:2019wpq,Wang:2017vwr}. In fact, all the mentioned results obtained for quantum field theories with boundary concern TFT: 3D Chern-Simons and BF theories, the latter being defined on any spacetime dimensions  \cite{Birmingham:1991ty, Horowitz:1989ng, Karlhede:1989hz, Blasi:2005vf,Blasi:2005bk}. What is still lacking is the study of a  physical, realistic, hence entirely non-topological, theory  in 4D, defined on a half-space. The first example which comes to mind of such a theory is of course 4D Maxwell theory of electromagnetism, and it is intriguing to investigate the role of the boundary in this case: which are the edge DOF? are there conserved currents, like in the topological cases? do they form an algebra? of which type? is there a 3D holographic counterpart of 4D Maxwell theory? is this unique, or more theories can be found on the 3D boundary, which are holographically compatible with the bulk theory? These questions motivated the present work, which is organised as follows. In Section 2 the boundary is introduced in 4D Maxwell theory. From the gauge fixed action, the equations of motion are derived, which yield the boundary conditions and the Ward identities, which are broken by the presence of the boundary. Then, from the Ward identities, equations are derived  which must be satisfied by  the two-point Green functions, $i.e.$ by the propagators, which, on-shell, are recognized to form a Ka\c{c}-Moody algebra. In addition, from the breakings of the Ward identities, the  DOF on the boundary are identified, as well as the symmetries which leave invariant their definition. It turns out that the symmetries are of the gauge type. In Section 3 the induced 3D theory is found in the following way. The Ka\c{c}-Moody algebra is interpreted as equal time commutators of canonical variables, and the possible Lagrangians which yield these commutation relations are considered, with a number of constraints amongst which is the gauge invariance. In Section 4 the holographic contact is performed. The correspondence between the bulk and the boundary theories is realized through a match between the equations of motion of the 3D theory and the boundary conditions found for the 4D bulk theory. There are two non equivalent ways to realize the contact, which remarkably land on the same 3D action. The energy momentum tensor is computed in Section 5, and, by imposing that its 00-component $i.e.$ the energy density, is positive, we determine the coefficients of the 3D theory, so that we may propose the 3D holographic counterpart of 4D Maxwell theory. Our results are summarized and discussed in the concluding Section 6. Appendices deal with specific analysis, namely in Appendix A we compute and discuss the propagators of the 3D theory, while in Appendix B we make some observations concerning symmetries of the bulk and boundary theory.
\\

In this paper we adopt the Minkowskian metric $\eta_{\mu\nu}=\mbox{diag}(-1,1,1,1)$. Our notations concerning indices, Levi-Civita tensors and coordinates  are as follows 
\begin{empheq}{align}
	\mu,\nu,\rho...&=\{0,1,2,3\}\nonumber\\
	\al,\be,\ga...&=\{0,1,2\}\label{eq:ind-d-1}\\
	i,j,k...&=\{1,2\}\ .\nonumber
\end{empheq}
\begin{equation}
\epsilon^{\al\beta\gamma}\equiv\epsilon^{\al\beta\gamma 3}\ .
\end{equation}
\begin{empheq}{align}
\mbox{4D bulk coordinates}: &\  x_\mu=(x_0,x_1,x_2,x_3)\\
\mbox{3D boundary $x_3=0$ coordinates }: &\		X_\al=(x_0,x_1,x_2)\ .
\end{empheq}

\section{The model: bulk and boundary}

\subsection{The action}  

The Minkowskian 4D Maxwell theory can be confined in the half-spacetime $x_3\geq 0$ by means of the introduction in the action of the Heaviside step function $\theta(x_3)$
		\begin{equation}\label{eq:Smax}
		S_M=-\frac{\ka}{4}\int d^4x\ \theta(x_3)\; F_{\mu\nu}F^{\mu\nu}\ ,
		\end{equation}
where $F_{\mu\nu}=\D_\mu A_\nu-\D_\nu A_\mu$ is the electromagnetic field strength, and $A_\mu(x)$ is the gauge field, with canonical mass dimension  $[A]=1$. In \eqref{eq:Smax}  $\ka>0$ is a constant which must be positive in order to have a positive-definite energy density. Maxwell theory, being a free field theory, does not display a coupling constant, which can always be reabsorbed by redefining the gauge field $A_\mu(x)$. Nonetheless, we do not normalize $\kappa$ to one, in order to be able to identify at any time the role played by the bulk action in the physics on the boundary.

The gauge fixing term		
\begin{equation}\label{eq:gauge-inv}
S_{gf}=\int d^4x\ \theta(x_3)\; bA_3
\end{equation}
implements, through the Lagrange multiplier field $b(x)$ \cite{Nakanishi:1966zz,Lautrup:1967zz}, the axial gauge condition
\begin{equation}
		A_3(x)=0\ .
\label{gaugecond}\end{equation}	
On the boundary $x_3=0$, the fields and their $\partial_3$-derivatives must be treated as independent fields \cite{Karabali:2015epa, Maggiore:2019wie}. To highlight this fact, we adopt the following notation:
\begin{equation}
		\At_\al(X)\equiv\left.\D_3A_\alpha\right|_{x_3=0}\ ,
\end{equation}	
whose mass dimension is $[\At]=2$. Therefore we must introduce another term in the action, coupling these two independent fields, $A_\mu(x)$ and, on the boundary, $\At_\al(X)$, to the external sources $J^\mu(x)$ and $\tilde J^\al(X)$ respectively:
		\begin{equation}
		S_J=\int d^4x\ \left( \theta(x_3)J^\al A_\al+\delta(x_3)\tilde{J}^\alpha \At_\al \right)\ .
		\end{equation}
The existence of the boundary requires an additional  contribution to the action~:
\begin{equation}\label{eq:Sbd}
				S_{bd}=\int d^4x\ \delta(x_3)\;
				\left(a^{\alpha\beta}A_\alpha A_\beta+b^{\al\be\gamma}\D_\al A_\be A_\gamma+c^{\al\be}\At_\al A_\be\right)\ ,
\end{equation}
where 
\begin{equation}
		a^{\al\be}=a^{\be\al},\quad b^{\al\be\gamma}=-b^{\al\gamma\be},\quad c^{\al\be}
\label{bdpar}\end{equation}
are constant matrices, with mass dimensions $[a^{\al\be}]=1,\ [b^{\al\be\gamma}]=[c^{\al\be}]=0$.
Such a lower dimensional term is tightly related to the presence of the boundary, and it must be present, wether the bulk action is gauge invariant, like in Maxwell case, or not, like in TFT, as a kind of counterterm, in a way similar to the Gibbons-Hawking term of General Relativity. The boundary term must only satisfy the general requirements of power counting, locality, and residual 3D Lorentz invariance. Gauge invariance must not be required on it: if we did it in TFT, we would have not recovered the boundary dynamics which characterizes those models. The total action, consisting in bulk term, gauge fixing, external sources and boundary contribution, finally is
\begin{equation}\label{eq:Stot}
		S_{tot}=S_M+S_{gf}+S_J+S_{bd}\ .
	\end{equation}

\subsection{Boundary conditions}\label{sec:Bd-cond}

From the action $S_{tot}$ \eqref{eq:Stot} we get the Equations Of Motion (EOM)
\begin{eqnarray}
\frac{\delta S_{tot}}{\delta A_\gamma(x)} &=&
\theta(x_3)[\ka\D_\mu F^{\mu\gamma}+J^\gamma] \nonumber \\
&& +\delta(x_3)[\ka F^{3\gamma}
			+2a^{\al\gamma}A_\al+2b^{\al\be\gamma}\D_\al A_\be+c^{\al\gamma}\At_\al]=0\label{eq:eom-max-bd1}\\
\frac{\delta S_{tot}}{\delta \At_\gamma(x)}&=&
-\ka\theta(x_3)F^{3\gamma}+\delta(x_3)[\tilde{J}^\gamma+c^{\gamma\al}A_\al]=0\ ,\label{eq:eom-max-bd2}
\end{eqnarray}
from which, by acting with the operator $\lim_{\epsilon\to0}\int_0^\epsilon dx_3$, and then going on-shell $\tilde{J}=0$, we derive the Boundary Conditions (BC)
\begin{empheq}{align}
		&\left. \ka\At^\gamma+2a^{\al\gamma}A_\al+2b^{\al\be\gamma}\D_\al A_\be+c^{\al\gamma}\At_\al\;  \right|_{x_3=0}=0\label{eq:bdc1}\\
		&\left. c^{\al\be}A_\be\; \right |_{x_3=0}=0\label{eq:bdc2}\ .
\end{empheq}
As it can be seen, this corresponds to putting equal to zero the $\delta(x_3)$ term of the EOM.
It is interesting to remark the analogy with the ``MIT bag model'' \cite{Chodos:1974je,Chodos:1974pn,DeGrand:1975cf,Johnson:1975zp}, which is one of the most successful phenomenological models for quark confinement. In this model, it is simply assumed that the quarks are confined to a spherical region of space (the ``bag''), with a radius $r = a$, and $V(r) = 0$ for $r < a$. Hence, the quark is treated as a free particle inside the region $r < a$, but is subject to boundary conditions (MIT bag model boundary conditions) at $r = a$ that realize    the confinement. This mechanism is obtained by means of the introduction in the action of a theta function, like we did in \eqref{eq:Smax} and of a boundary term proportional to a delta function, in close analogy with \eqref{eq:Sbd}. Consequently, the EOM have the same structure as \eqref{eq:eom-max-bd1} and \eqref{eq:eom-max-bd2}, $i.e.$ they are formed by two parts (theta and delta dependent). The MIT bag model boundary conditions are realized by putting equal to zero the delta dependent part, exactly as we did to obtain \eqref{eq:bdc1} and \eqref{eq:bdc2}. Moreover, instead of introducing theta functions by hand in the action, MIT boundary conditions can be induced dynamically, as discussed in \cite{Guendelman:2015qva}. This remark suggests a possible application of the method presented in this paper to the MIT bag model.\footnote{We thank the Referee for pointing out this analogy.}

\subsection{Ward identities}\label{sec:wi-max}

The EOM \eqref{eq:eom-max-bd1} and \eqref{eq:eom-max-bd2} give rise to the Ward identities, crucial for what follows. From \eqref{eq:eom-max-bd1}, we have

\begin{equation}
			\begin{split}
			\int^{+\infty}_{0}dx_3\; \D^\gamma J_\gamma&
			=-\ka\int^{+\infty}_{-\infty}dx_3\ \theta(x_3)\; \D^\gamma\D^\mu F_{\mu\gamma}\\
			&=-\ka\int^{+\infty}_{-\infty}dx_3\ \theta(x_3)\; (\D^\gamma\D^\be F_{\be\gamma}+\D^\gamma\D^3 F_{3\gamma})\\
			&=\ \ka\int^{+\infty}_{-\infty}dx_3\ \delta(x_3)\; \D^\gamma F_{3\gamma}\\
			&=\ka\; \left.\D^\gamma\At_\gamma\right|_{x_3=0}\ ,
			\label{eq:wi1}\end{split}
		\end{equation}
where we used 		
		\begin{equation}
		\D_3\theta(x_3)=\delta(x_3)\ .
		\end{equation}
Analogously, from \eqref{eq:eom-max-bd2} we find
\begin{equation}\label{eq:wi2}
		\D^\gamma\tilde{J}_\gamma|_{x_3=0}=-\ka\; \left.\D^\gamma A_\gamma\right|_{x_3=0}\ .
		\end{equation}
Notice that the Ward identity \eqref{eq:wi2}, differently from \eqref{eq:wi1}, is local and not integrated. Remark also that both the Ward identities \eqref{eq:wi1} and \eqref{eq:wi2}  are broken, because of the presence of the boundary, by a linear term at their r.h.s. Such Ward identities are known \cite{mack,Becchi:1988nh} to imply conservation laws. In fact,} at vanishing external sources $\tilde J=J=0$, $i.e.$ going on shell, we find
		\begin{empheq}{align}
		\D^\al \At_\al|_{x_3=0}&=0\label{subeq:cc1}\\
		\D^\al A_{\al}|_{x_3=0}&=0\ ,\label{subeq:cc2}
		\end{empheq}
which show the existence of a couple of conserved currents on the 3D edge of 4D Maxwell theory.
		
\subsection{Algebra}\label{sec:Alg}

Once the generating functional of connected Green functions $Z_c[J,\tilde{J}]$ has been defined in the usual way
\begin{equation}
e^{iZ_c[J,\tilde{J}]}=\int DA D\At Db\; e^{iS_{tot}[A,\At,b;J,\tilde{J}]}\ ,
\end{equation}
the following relations hold
\begin{eqnarray}
\left.\frac{\delta Z_c[J]}{\delta J^\al(x)}\right|_{J=0} &=& A_\al(x)     \label{eq:A} \\
\frac{\delta^{(2)} Z_c[J]}{\delta J^\al(x)\delta J^\be(x')}\bigg|_{J=0}&=&i\langle T(A_\al(x)A_\be(x')\rangle\ ,\label{eq:TprodA}
\end{eqnarray}
where the time-ordered product is defined as
\begin{equation}
\langle T(A_\al(x)A_\be(x'))\rangle \equiv
\theta(x_0-x'_0) \langle A_\al(x)A_\be(x')\rangle +
\theta(x'_0-x_0) \langle A_\be(x')A_\al(x)\rangle\ .
\end{equation}
Differentiating the first Ward identity \eqref{eq:wi1} with respect to $J_\beta(x')$ and then going on-shell  $J=\tilde{J}=0$, we have
\begin{eqnarray}
\D^\be\delta^{(3)}(X-X')&=&
i\ka\; \D^\al \langle T(\At_\al(X)A^\be(X')) \rangle \nonumber\\
&=& 
i\ka\; [\At_0(X),A^\be(X')]\delta(x_0-x'_0)+
			i\ka\; \langle T(\D^\al\At_\al(X)A^\be(X'))\rangle\ ,\label{2.32}
\end{eqnarray}
where we used
\begin{equation}
		\frac{\delta J_\al(x)}{\delta J_\be(x')}=\delta^\be_{\ \al}\delta^{(4)}(x-x')\ .
\end{equation}
Choosing $\be=0$ in \eqref{2.32}, and remembering that, on-shell, \eqref{subeq:cc1} holds, the second term on the r.h.s. of \eqref{2.32} vanishes, and we get
		\begin{equation}
		\delta(x_0-x'_0)[\At_0(X),A_0(X')]=\frac{i}{\ka}\D^0\delta^{(3)}(X-X')\ .
		\label{2.34}\end{equation}
By integrating with respect to $x_0$ both sides of \eqref{2.34}, we are left with the equal time commutator
\begin{equation}\label{eq:a0at0}
		[\At_0(X),A_0(X')]_{_{x_0=x'_0}}=0\ .
\end{equation}
If, instead, $\be=i$ in \eqref{2.32}, then $\delta(x_0-x'_0)$ can be factorized, and we find		
\begin{equation}
			[\At_0(X),A_i(X')]=-\frac{i}{\ka}\D_i\delta^{(2)}(X-X')\ .
		\end{equation}
By differentiating  the first Ward identity \eqref{eq:wi1} with respect to $\tilde{J}^\be(x')$, we get
\begin{equation}
			\begin{split}
			0&=\D^\al\frac{\delta^{(2)} Z_c[J,\tilde{J}]}{\delta \tilde{J}^\al(X)\delta\tilde{J}^\be(X')}\bigg|_{J=\tilde{J}=0}\\
			&=\D^\al \langle T(\At_\al(X)\At_\be(X'))\rangle\\
		&=[\At_0(X),\At_\be(X')]\delta(x_0-x'_0)+\langle T(\D^\al\At_\al(X)\At_\be(X'))\rangle\ ,
			\end{split}	
		\end{equation}
which, using again the current conservation on the edge \eqref{subeq:cc1}, leads to the equal time commutator
\begin{equation}
		[\At_0(X),\At_\al(X')]_{_{x_0=x'_0}}=0\ .
\end{equation}
We can extract similar informations from the second, local, Ward identity \eqref{eq:wi2}, which, differentiated with respect to $\tilde{J}_\be(x')$ and put on-shell
\begin{equation}
\begin{split}
			\D^\al\frac{\delta \tilde{J}_\al(X)}{\delta\tilde{J}_\be(X')}\bigg|_{J=\tilde{J}=0}&=
	-\ka\D^\al\frac{\delta^{(2)} Z_c[J,\tilde{J}]}{\delta J^\al(X)\delta\tilde{J}_\be(X')}\bigg|_{J=\tilde{J}=0}\ ,
\end{split}	
\end{equation}
gives
\begin{equation}
\begin{split}
			\D^\be\delta^{(3)}(X-X')&=-i\ka\D^\al\langle T(A_\al(X)\At^\be(X'))\rangle\\
		&=-i\ka[A_0(X),\At^\be(X')]\delta(x_0-x'_0)-i\ka\langle T(\D^\al A_\al(X)\At^\be(X'))\rangle\ .
			\end{split}
\label{2.40}\end{equation}
For $\be=0$ we get the same result as in \eqref{eq:a0at0}, as a check of the coherence of our way to proceed. Putting $\be=i$ in \eqref{2.40}, and using \eqref{subeq:cc2}, we have, again at equal time:
\begin{equation}
		[A_0(X),\At_i(X')]=\frac{i}{\ka}\D_i\delta^{(2)}(X-X')\ .
\end{equation}
Finally, differentiating the local Ward identity \eqref{eq:wi2} with respect to $J^{\be}(x')$, we get
\begin{equation}
			\begin{split}
			0&=\D^\al\frac{\delta^{(2)} Z_c[J,\tilde{J}]}{\delta J^\al(X)\delta J^\be(X')}\bigg|_{J=\tilde{J}=0}\\
			&=\D^\al\langle T(A_\al(X)A_\be(X'))\rangle\\
		&=[A_0(X),A_\be(X')]\delta(x_0-x'_0)+\langle T(\D^\al A_\al(X)A_\be(X'))\rangle\ ,
			\end{split}
\end{equation}
finding the equal time commutator 
\begin{equation}
		[A_0(X),A_\be(X')]_{_{x_0=x'_0}}=0\ .
		\end{equation}
Summarizing, from the Ward identities \eqref{eq:wi1} and \eqref{eq:wi2}, broken by the presence of the boundary $x_3=0$, we get the following equal time commutators for the conserved currents $A_\al(X)$ and $\At_\al(X)$
\begin{empheq}{align}			 
		[\At_0(X),A_i(X')]&=-\frac{i}{\ka}\D_i\delta^{(2)}(X-X')\label{eq:[At0,Ai]}\\
		[A_0(X),\At_i(X')]&=\frac{i}{\ka}\D_i\delta^{(2)}(X-X')\label{eq:[A0,Ati]}\\
		[\At_0(X),\At_\al(X')]&=[A_0(X),A_\al(X')]=[\At_0(X),A_0(X')]=0\ .\label{eq:[]0}
		\end{empheq}
In order to identify the correct DOF on the 3D boundary, it is convenient to introduce a field $B_\al(X)$ defined by the linear transformations
\begin{equation}
\begin{split}
B_0\equiv &\ \mu A_0+\nu\At_0 \\
B_i\equiv&\ \rho A_i+\sigma\At_i	\ ,		
\label{defB}\end{split}
\end{equation}
where $\mu$, $\nu$, $\rho$ and $\sigma$ are constant parameters, which will be set later at our convenience, with mass dimensions constrained by the request of dimensional homogeneity of \eqref{defB}
\begin{equation}
			[\mu]=[\nu]+1\ ,\qquad[\rho]=[\sigma]+1\ ,\qquad[\mu]=[\rho]\quad\mathrm{and}\quad[\nu]=[\sigma]\ .
\label{dimpar}\end{equation}
In terms of $B_\al(X)$, the algebra \eqref{eq:[At0,Ai]}-\eqref{eq:[]0} reduces to the only nonvanishing commutator
\begin{equation}
[B_0(X),B_i(X')] =\;i\;\frac{\mu\sigma-\nu\rho}{\ka}\;\D_i\delta^{(2)}(X-X')\ ,
\label{eq:B0,Bi}
\end{equation}
which describes an abelian Ka\c{c}-Moody algebra whose central charge is proportional to the inverse of the Maxwell coupling $\kappa$:
\begin{equation}
	\frac{1}{\kat}\equiv\frac{\mu\sigma-\nu\rho}{\ka}\ .\label{eq:condiz1}
\end{equation}
We therefore recover for 4D Maxwell theory with boundary a property peculiar to TFT \cite{Blasi:1990pf,Blasi:1990bk,Emery:1991tf}. A comment is here in order: Conformal Field Theories in two and more dimensions are classified in terms of the central charges of their Ka\c{c}-Moody algebras, 
which should be positive, for the unitarity of the theory \cite{mack,Becchi:1988nh}. Remembering that the Maxwell coupling constant $\kappa$ is positive, we thus have the constraint
\begin{equation}
	{\mu\sigma-\nu\rho}>0\ .\label{eq:condiz2}
\end{equation}
Interestingly enough, we shall see in Section 5 that this requirement is strictly related to the positivity of the energy density of the 3D theory we shall find on the boundary. Moreover, we have a physical interpretation of the parameters appearing in \eqref{defB}: each set $(\mu,\nu,\rho,\sigma)$ respecting \eqref{eq:condiz2} corresponds to a different central charge, hence to a different Conformal Field Theory. This is an important novelty with respect to TFT, where, instead, there is a bijection between bulk coupling constants and central charges, which in this case is realized only if
\begin{equation}
	{\mu\sigma-\nu\rho}=1\ .\label{2.42}
\end{equation}

\subsection{Boundary dynamics}\label{sec:bd-dyn}

The 3D current conservation relations \eqref{subeq:cc1} and \eqref{subeq:cc2} can be solved by
\begin{empheq}{align}			 
		&\At_\al(X)=\epsilon_{\al\be\gamma}\D^\be \tilde{\xi}^\gamma(X)\label{eq:xit}\\
		&A_{\al}(X)=\epsilon_{\al\be\gamma}\D^\be\xi^\gamma(X)\ ,\label{eq:xi}
		\end{empheq}
where the fields $\tilde{\xi}_\al(X)$ and $\xi_\al(X)$ have canonical dimensions 
		\begin{equation}
		[\tilde{\xi}]=1\quad\mathrm{and}\quad[\xi]=0\ .\label{eq:dim-xi}
		\end{equation}	
Consequently, we have 
\begin{empheq}{align}
B_0=&\epsilon_{0ij}\D^i(\mu\xi^j+\nu\xit^j)=\epsilon_{0ij}\D^i\lambda^j\label{eq:B0-bd}\\
		B_i=&\epsilon_{i\al\be}\D^\al(\rho\xi^\be+\sigma\xit^\be)
		=\epsilon_{i\al\be}\D^\al\tilde{\lambda}^\be\label{eq:Bi-bd}\ ,			 
		\end{empheq}
where we defined
		\begin{empheq}{align}			 
			\la_\al&\equiv\mu\xi_\al+\nu\xit_\al\label{eq:la-bd}\\
			\lat_\al&\equiv\rho\xi_\al+\sigma\xit_\al\label{eq:lat-bd}\ .		 
		\end{empheq}
The equations \eqref{eq:la-bd} and \eqref{eq:lat-bd} define the 3D vector fields  $\la_\al(X)$ and $\lat_\al(X)$ which, as we shall show in what follows, are the dynamical variables in terms of which the 3D theory induced on the boundary of 4D Maxwell theory will be constructed.  Notice that the defining relations \eqref{eq:B0-bd} and \eqref{eq:Bi-bd} are left invariant under the transformations 
\begin{empheq}{align}				 
				\lambda_\al\quad\to&\quad\lambda_\al+\D_\al\Lambda\label{eq:dla}\\
				\lat_\be\quad\to&\quad\lat_\be+\D_\be\tilde{\Lambda}\ ,\label{eq:dlat}				 
\end{empheq}
where $\La(X)$ and  $\Lat(X)$ are local gauge parameters. For what concerns the canonical mass dimensions of $\la_\al(X)$ and $\lat_\al(X)$, the standard possibilities for 3D gauge fields are
\begin{empheq}{align}
		a)\quad&[\la]=[\lat]=1\label{eq:dim=1}\\
		b)\quad&[\la]=[\lat]=\frac{1}{2}\ .\label{eq:dim=1/2}
\end{empheq}
The first choice involves, for instance, topological Chern-Simons theory, or any 3D theory involving one derivative only in its quadratic term. The second possibility, instead, is mandatory for 3D gauge field theories with two derivatives, like for instance Maxwell theory, possibly coupled with a topological Chern-Simons term by means of a massive parameter, like topologically massive 3D Maxwell-Chern-Simons theory \cite{Deser:1982vy}. It is interesting to remark that, in principle, the definition of $B_\al(X)$ given by \eqref{defB} allows both possibilities, which is a nontrivial fact, because of the two dimensional constraints on the fields $\xi_\al(X)$ and $\xit_\al(X)$  in \eqref{eq:dim-xi} and on the parameters $\{\mu,\nu,\rho,\sigma\}$ in \eqref{dimpar}. Indeed
\begin{itemize}
\item[$a$)] by choosing 
			\begin{equation}
			[\mu]=[\rho]=1\quad\mathrm{and}\quad[\nu]=[\sigma]=0
			\label{2.63}\end{equation}
then
			\begin{equation}
			[\la]=[\lat]=1\ ,
			\label{2.64}\end{equation}
to which corresponds $[\kat]=-1$\ ;
\item[$b$)] if instead
			\begin{equation}
			[\mu]=[\rho]=\frac{1}{2}\quad\mathrm{and}\quad[\nu]=[\sigma]=-\frac{1}{2}
			\label{2.65}\end{equation}
we have 
			\begin{equation}
			[\la]=[\lat]=\frac{1}{2}\ .
			\label{2.66}\end{equation}
In this case the central charge of the Ka\c{c}-Moody algebra \eqref{eq:B0,Bi} formed by the fields $B_0$ and $B_i$ has, like the 4D Maxwell coupling $\ka$, vanishing dimension: $[\kat]=0.$
\end{itemize}
In this paper we shall study both possibilities. 

\section{Induced 3D theory}

In the previous Section we identified the 3D DOF  induced on the boundary of 4D Maxwell theory as the two vector fields $\lambda_\al(X)$ and $\tilde\lambda_\al(X)$ defined by \eqref{eq:B0-bd} and \eqref{eq:Bi-bd}. This same definition is left invariant by the transformations \eqref{eq:dla} and \eqref{eq:dlat}. We may therefore claim that the 3D theory induced on the boundary of 4D Maxwell theory should be a gauge theory of two, possibly coupled, gauge fields, which must satisfy the following three constraints: 
\begin{description}
\item[1:] invariance under the gauge transformations
\begin{empheq}{align}
		&\delta_1\la_\al(X)=\D_\al\La(X)\label{eq:gt-1}\\
		&\delta_2\lat_\al(X)=\D_\al\Lat(X)\label{eq:gt-2}\ ;
\end{empheq}
\item[2:] compatibility with the equal time Ka\c{c}-Moody algebra \eqref{eq:B0,Bi}\ ; 
\item[3:] compatibility with the BC \eqref{eq:bdc1} and \eqref{eq:bdc2}\ .
\end{description}
For what concerns the constraint {\bf 2}, the equal time Ka\c{c}-Moody algebra \eqref{eq:B0,Bi}, written
in terms of the boundary fields $\lambda_\al(X)$ and $\tilde\lambda_\al(X)$, becomes
\begin{equation}
		[\epsilon_{0ij}\la^j(X),\epsilon^{k\al\be}\D'_\al\lat_\be(X')]=
		\frac{{i}}{\kat}\delta_i^k\delta^{(2)}(X-X')\ ,
		\end{equation}
which, in the temporal gauge 
\begin{equation}
		\la_0=\lat_0=0\ ,
\label{tempgauge}\end{equation}
reads
\begin{equation}\label{eq:alg-la-km}
		[\epsilon_{0ij}\la^j(X),-\epsilon^{0kl}\D'_0\lat_l(X')]=\frac{{i}}{\kat}\delta_i^k\delta^{(2)}(X-X')\ .\end{equation}
The key observation is to recognize in this algebra the canonical commutation relations
\begin{equation}\label{eq:alg-la-pq}
		[q_i(X),p^j(X')]={i}\delta_i^j\delta^{(2)}(X-X')\ ,
\end{equation}
once the canonical variables have been identified as
\begin{empheq}{align}			 
			q_i=&\quad\kat\epsilon_{0ij}\lambda^j \label{eq:q}\\
			p^i=&-\epsilon^{0ij}\D_0\lat_j\ ,\label{eq:p}			 
\end{empheq}
where the conjugate momentum $p^i$ is defined as
\begin{equation}
p^i=\frac{\D\Lag}{\D\dot q_i}\ .
\label{defp}\end{equation}
The gauge invariant action satisfying the constraint {\bf 1} should depend on the fields $\lambda_\al(X)$ and $\tilde\lambda_\al(X)$ in a way 
to preserve the relation \eqref{defp} between the canonical variables $q_i(X)$ \eqref{eq:q} and $p^i(X)$ \eqref{eq:p}. 
Finally, the action satisfying the first two constraints should display EOM compatible with the BC of the bulk theory \eqref{eq:bdc1} and \eqref{eq:bdc2}. We are now ready to analyze the two possible cases \eqref{eq:dim=1} and \eqref{eq:dim=1/2}.

\subsection{Case $a$: $[\la]=[\lat]=1$ }

The canonical mass dimensions of the two vector fields of the 3D theory are both set to one. This can be realized through the choice \eqref{2.63}. We are looking for the most general quadratic Lagrangian 
$\Lag_{(a)}[\lambda,\lat]$ respecting the power counting \eqref{eq:dim=1} and whose action is invariant under the gauge transformations \eqref{eq:gt-1} and \eqref{eq:gt-2} (constraint {\bf 1.})~:
\begin{equation}
		\delta_{1,2}S_{(a)}[\lambda,\lat]=\delta_{1,2}\int d^3X\ \Lag_{(a)}[\lambda,\lat]=0\ .
\end{equation}
It is immediate to verify that the result is 
\begin{equation}
		\Lag_{(a)}[\lambda,\lat]=k_1\epsilon^{\al\be\gamma}\D_\al\la_\be\la_\gamma+k_2\epsilon^{\al\be\gamma}\D_\al\la_\be\lat_\gamma+k_3\epsilon^{\al\be\gamma}\D_\al\lat_\be\lat_\gamma\ ,
		\label{casei}\end{equation}		
where the coefficients $k_i$ have vanishing mass dimensions
\begin{equation}
		[k_i]=0\ .
\end{equation}
We find that the Lagrangian \eqref{casei} contains only topological terms, of the Chern-Simons and BF type. Let us consider now the constraint {\bf 2}, which we showed to be equivalent to the definition of the canonical variables \eqref{eq:q} and \eqref{eq:p}, related by \eqref{defp}
\begin{equation}\label{eq:dL/dq=p'}
		\frac{\D\Lag_{(a)}}{\D\dot q_i}=p^i=-\epsilon^{0ij}\D_0\lat_j\ ,
\end{equation}
where we used the temporal gauge choice \eqref{tempgauge}. The l.h.s. of this translates in
\begin{equation}\label{Li}
			\begin{split}
			\frac{\D\Lag_{(a)}}{\D\dot q_i}&=\frac{1}{\kat}\epsilon^{0ij}\frac{\D\Lag_{(a)}}{\D(\D^0\la^j)}\\
			&=\frac{1}{\kat}\epsilon^{0ij}\bigl[k_1\epsilon_{0jk}\la^k+k_2\epsilon_{0jk}\lat^k\bigr]\\
			&={-}\frac{1}{\kat}\bigl[k_1\la^i+k_2\lat^i\bigr]\ .
			\end{split}
\end{equation}
Comparing \eqref{eq:dL/dq=p'} with \eqref{Li}, it appears that it is not possible to set the parameters $k_i$ in such a way that the relation \eqref{eq:dL/dq=p'} is verified.\\

We therefore proved a first nontrivial result: the 4D Maxwell theory cannot induce on its 3D boundary a purely TFT.

\subsection{Case $b$: $[\la]=[\lat]=\frac{1}{2}$ }

In this case, it is easy to show that the most general quadratic Lagrangian $\Lag_{(b)}[\lambda,\lat]$ com\-pa\-tible with the dimensional assignments \eqref{eq:dim=1/2} and whose action is invariant under the gauge transformations \eqref{eq:gt-1} and \eqref{eq:gt-2}, is the following
\begin{equation}
			\begin{split}
			\Lag_{(b)}=&k_1G_{\al\be}G^{\al\be}+k_2G_{\al\be}\Gt^{\al\be}+k_3\Gt_{\al\be}\Gt^{\al\be}\\
			&+m_1\epsilon^{\al\be\gamma}\D_\al\la_\be\la_\gamma+m_2\epsilon^{\al\be\gamma}\D_\al\la_\be\lat_\gamma+m_3\epsilon^{\al\be\gamma}\D_\al\lat_\be\lat_\gamma\ ,
			\end{split}
		\label{Lii}\end{equation}
where we defined the field strengths for the 3D gauge fields $\la_\al(X)$ and $\lat_\al(X)$
\begin{empheq}{align}
		G_{\al\be}&\equiv\D_\al\la_\be-\D_\be\la_\al\\
		\Gt_{\al\be}&\equiv\D_\al\lat_\be-\D_\be\lat_\al\ ,
\end{empheq}		
and the coefficients have mass dimensions
\begin{equation}
		[k_i]=0\ ;\ [m_i]=1\ .
\end{equation}	
The theory described by the Lagrangian \eqref{Lii} is not purely topological, like \eqref{casei}, but it contains topological terms, of both the Chern-Simons and BF type. Let us proceed now to check whether the constraint {\bf 2}, concerning the identification of the canonical variables $q_i(X)$ and $p^i(X)$, is fulfilled, which means 
		\begin{equation}\label{eq:dL/dq=p'2}\frac{\D\Lag_{(b)}}{\D\dot q_i}=p^i=-\epsilon^{0ij}\D_0\lat_j\ .
		\end{equation}
We have 
\begin{equation}\label{eq:dl1-dq}
			\begin{split}
			\frac{\D\Lag_{(b)}}{\D\dot q_i}&=\frac{1}{\kat}\epsilon^{0ij}\frac{\D\Lag_{(b)}}{\D(\D^0\la^j)}\\
			&=\frac{1}{\kat}\epsilon^{0ij}\bigl[4k_1\D_0\la_j+2k_2\D_0\lat_j+m_1\epsilon_{0jk}\la^k+m_2\epsilon_{0jk}\lat^k\bigr]\\
			&=\frac{1}{\kat}\bigl[4k_1\epsilon^{0ij}\D_0\la_j+2k_2\epsilon^{0ij}\D_0\lat_j{-}m_1\la^i{-}m_2\lat^i\bigr]\ .
			\end{split}
\end{equation}		
We see that the above expression matches \eqref{eq:dL/dq=p'2} if
\begin{equation}
		k_1=m_1=m_2=0\qquad\mathrm{and}\qquad k_2=-\frac{\kat}{2}\ .
\end{equation}
Therefore, a possible candidate for Case $b$ exists, which is represented by the 3D action
\begin{equation}\label{eq:S-gen}
		S_{3D}\equiv\int d^3X\ \Lag_{(b)}=\int d^3X\biggl(-\frac{\kat}{2}G_{\al\be}\Gt^{\al\be}+k_3\Gt_{\al\be}\Gt^{\al\be}+m_3\epsilon^{\al\be\gamma}\D_\al\lat_\be\lat_\gamma\biggr)\ .
		\end{equation}
We can summarize what we found so far as follows: the 4D Maxwell theory \eqref{eq:Smax}, after the definition of the fields \eqref{defB}, shows on its planar boundary the algebra of the Ka\c{c}-Moody type \eqref{eq:B0,Bi}, which can be interpreted as a canonical commutation relation \eqref{eq:alg-la-pq}. The boundary DOF are identified as two vector fields $\lambda_\alpha(X)$ and $\tilde\lambda_\alpha(X)$, which must have mass dimensions 1/2, no other choices being possible. The gauge invariances \eqref{eq:gt-1} and \eqref{eq:gt-2} of the 3D theory are not a request, but rather a consequence of the definitions \eqref{eq:B0-bd} and \eqref{eq:Bi-bd}. We showed that the gauge invariant 3D action \eqref{eq:S-gen} respects the relation \eqref{defp} between canonical variables. What is left to implement is the constraint {\bf 3}, concerning the compatibility of this new 3D theory with the BC \eqref{eq:bdc1} and \eqref{eq:bdc2} of the 4D bulk action. This nontrivial task, which we call holographic contact, will be achieved in the next Section.

\section{Holographic contact}

The 3D theory \eqref{eq:S-gen} can be seen as the holographic counterpart of the 4D bulk Maxwell theory \eqref{eq:Smax} once the constraint {\bf 3} is fulfilled. This result is obtained by matching the BC \eqref{eq:bdc1} and \eqref{eq:bdc2} of the 4D theory  with the EOM obtained from the 3D action \eqref{eq:S-gen}, which are
\begin{empheq}{align}
		\frac{\delta S_{3D}}{\delta\la_\gamma}=&\ \kat\D_\al\Gt^{\al\gamma}=0\label{eq:eom-la}\\
		\frac{\delta S_{3D}}{\delta\lat_\gamma}=&\ \kat\D_\al G^{\al\gamma}-4k_3\D_\al\Gt^{\al\gamma}+2m_3\epsilon^{\al\be\gamma}\D_\al\lat_\be=0\ .\label{eq:eom-lat}
\end{empheq}
The contact is made by relating the coefficients	$a^{\al\be}$, $b^{\al\be\gamma}$ and $c^{\al\be}$, appearing in the boundary term $S_{bd}$ \eqref{eq:Sbd} with $\kat$, $k_3$ and $m_3$, which are the parameters of the 3D action \eqref{eq:S-gen}. The EOM \eqref{eq:eom-la} and \eqref{eq:eom-lat} are written in terms of the fields $\lambda_\alpha(X)$ and $\tilde\lambda_\alpha(X)$, while the BC \eqref{eq:bdc1} and \eqref{eq:bdc2} depend on the 4D gauge field and its $\partial_3$-derivative on the boudary $x_3=0$: $\left.A_\alpha(x)\right|_{x_3=0}$ and $\left.\partial_3A_\alpha(x)\right|_{x_3=0}$. 
Therefore, as a preliminary step, we have to write the four equations involved in terms of the same fields, and the most convenient choice is to express everything in terms of $\xi_\alpha(X)$ \eqref{eq:xi} and $\tilde\xi_\alpha(X)$  \eqref{eq:xit}, which are related to $\lambda_\alpha(X)$ and $\tilde\lambda_\alpha(X)$ by means of \eqref{eq:la-bd} and \eqref{eq:lat-bd}, respectively. 
To do that, we write
\begin{equation}\label{eq:Gmunu}
		G_{\al\be}=\D_\al\la_\be-\D_\be\la_\al=(\delta^\eta_\al\delta^\gamma_\be-\delta^\gamma_\al\delta^\eta_\be)\D_\eta\la_\gamma=\epsilon_{\al\be\delta}\epsilon^{\eta\gamma\delta}\D_\eta\la_\gamma\ ,
\end{equation}
which we can use to rewrite the EOM \eqref{eq:eom-la} as
\begin{equation}
	0=\D_\al\Gt^{\al\be}=\epsilon^{\al\be\gamma}\D_\al\bigl(\epsilon_{\theta\delta\gamma}\D^\theta\lat^\delta\bigr)\ .
\end{equation}
In terms of $\xi_\al(X)$ and $\xit_\al(X)$ this translates into 
	\begin{equation}\label{eq:eom-xi}
	\epsilon_{\la\ka\al}\D^\la\bigl[\eta^{\al\be}\epsilon_{\gamma\delta\be}\D^\gamma(\rho\xi^\delta+\sigma\xit^\delta)\bigr]=0\ ,
	\end{equation}
where we used \eqref{eq:lat-bd}. In the same way, using \eqref{eq:Gmunu} in the EOM \eqref{eq:eom-lat}, we find
\begin{equation}\label{eq:eom-la-2}
			\begin{split}
			0&=\D_\al\bigl[\kat G^{\al\gamma}-4k_3\Gt^{\al\gamma}+2m_3\epsilon^{\al\be\gamma}\lat_\be\bigr]\\
			&=\epsilon^{\al\be\gamma}\D_\al\bigl[-\kat\ \epsilon_{\theta\delta\be}\D^\theta\la^\delta+4k_3\ \epsilon_{\theta\delta\be}\D^\theta\lat^\delta+2m_3\lat_\be\bigr]\ ,
			\end{split}
\end{equation}
which, in terms of $\xi_\al(X)$ and $\xit_\al(X)$, becomes
\begin{equation}\label{eq:eom-xi-2}
		\epsilon_{\la\al\theta}\D^\la\bigl\{\epsilon_{\be\ga\delta}\D^\gamma\bigl[(\kat\nu-4k_3\sigma)\eta^{\al\be}\xit^\delta+(\kat\mu-4k_3\rho)\eta^{\al\be}\xi^\delta\bigr]\bigr\}-
		2m_3\epsilon_{\la\al\theta}\D^\la(\rho\xi^\al+\sigma\xit^\al)=0\ .
\end{equation}
Hence, the EOM \eqref{eq:eom-la} and \eqref{eq:eom-lat}, written in terms of the boundary fields 
$\xi_\al(X)$ and $\xit_\al(X)$, are \eqref{eq:eom-xi} and \eqref{eq:eom-xi-2}, respectively. We are now able to compare them with the BC \eqref{eq:bdc1} and \eqref{eq:bdc2}, which, written in terms of the same variables, are~:
\begin{empheq}{align}			 
				&\epsilon_{\be\gamma\delta}\D^\gamma [(\ka\eta^{\al\be}+c^{\be\al})\tilde{\xi}^\delta+2(a^{\al\be}+b^{\kappa\be\al}\D_\kappa)\xi^\delta]=0\label{eq:bdc1'}\\
				&c^{\al\be}\epsilon_{\be\gamma\delta}\D^\gamma\xi^\delta=0\ .\label{eq:bdc2'}
\end{empheq}	
We observe that, since the mass dimensions of the EOM and of the BC differ, in order to compare them we will need to introduce massive coefficients and/or derivatives. We found that the various possibilities of contact eventually fall into two  inequivalent categories, which we schematically represent as follows:
\begin{itemize}
		\item[{\bf 1:}] 
		\begin{empheq}{align} 
		(\ref{eq:eom-xi})&\leftrightarrow c_1\; \mathrm{curl}(\ref{eq:bdc2'})\label{eq:cont3.1}\\
		(\ref{eq:eom-xi-2})&\leftrightarrow c_2\; (\ref{eq:bdc2'})+c_3\; \mathrm{curl}(\ref{eq:bdc1'})\label{eq:cont3.2}\ ,
		\end{empheq}
		\item[{\bf 2:}]
		\begin{empheq}{align} 
		(\ref{eq:eom-xi})&\leftrightarrow c_4\; \mathrm{curl}(\ref{eq:bdc1'})\label{eq:cont4.1}\\
		(\ref{eq:eom-xi-2})&\leftrightarrow c_5\;(\ref{eq:bdc1'})+c_6\;\mathrm{curl}(\ref{eq:bdc2'})\ ,\label{eq:cont4.2}
		\end{empheq}
\end{itemize} 
where $c_1$,  $c_2$,  $c_3$, $c_4$, $c_5$ and  $c_6$ are  parameters with the following mass dimensions
\begin{equation}
[c_1]= 1/2 \ ;\ [c_2]= 3/2 \ ;\ [c_3]= -1/2 \ ;\ [c_4]= -1/2 \ ;\ [c_5]=1/2\ ;\ [c_6]=1/2\ .
\end{equation}
In the above expressions, by ``curl(eq.)'' we mean the curl of the equation in parenthesis~:
	\begin{equation}
	\mbox{curl(eq.)} = \epsilon_{\al\be\ga}\D^\be(\mathrm{eq.})^\ga\ .
	\end{equation}
We proceed now to study the above two possibilities in details. We are mostly interested in finding out whether the two cases yield compatible 3D theories, and whether these theories are equivalent one to each other or not. 

\subsection{Case 1}\label{sec:c1}

We have to relate the EOM (\ref{eq:eom-xi}) and the BC (\ref{eq:bdc2'}) by means of 
\begin{equation}
	(\ref{eq:eom-xi})\leftrightarrow c_1\;\mathrm{curl}(\ref{eq:bdc2'})\ ,
\end{equation}
which can be obtained if 
\begin{empheq}{align}
		&\sigma=0\label{eq:sig0-3}\\
		&c^{\al\be}=\frac{\rho}{ c_1}\;\eta^{\al\be}\label{eq:c-ab1}\ .
\end{empheq}
As a consequence of (\ref{eq:sig0-3}), from \eqref{eq:condiz1} we get
\begin{equation}\label{eq:nu3}
		\nu={-}\frac{\ka}{\kat\rho}\ ,
\end{equation}
which allows us to write the EOM (\ref{eq:eom-xi-2}) as
\begin{equation}\label{eq:eom-3}
		-2\rho m_3\eta^{\ka\be}\epsilon_{\be\gamma\delta}\D^\gamma\xi^\delta+\epsilon^{\ka\la\al}\D_\la\bigl\{\epsilon^{\be\gamma\delta}\D_\gamma\bigl[{-}\tfrac{\ka}{\rho}\ \eta_{\al\be}\xit_\delta+2(\tfrac{\kat\mu}{2}-2\rho k_3)\eta_{\al\be}\xi_\delta\bigr]\bigr\}=0\ .
\end{equation}
The linear combination of the BC: 
\begin{equation}
		c_2\;(\ref{eq:bdc2'})+c_3\;\mathrm{curl}(\ref{eq:bdc1'})
\end{equation}
explicitly reads
		\begin{equation}\label{eq:bdc-3}
		c_2\;c^{\ka\be}\epsilon_{\be\gamma\delta}\D^\gamma\xi^\delta+c_3\;\epsilon^{\ka\la\al}\D_\la\bigl\{\epsilon^{\be\gamma\delta}\D_\gamma\bigl[(\ka\eta_{\al\be}+c_{\be\al})\xit_\delta+2a_{\al\be}\xi_\delta+2b_{\theta\be\al}\D^\theta\xi_\delta\bigr]\bigr\}=0\ .
		\end{equation}
The contact between 4D and 3D theories is achieved if \eqref{eq:eom-3} = \eqref{eq:bdc-3}, $i.e.$
		\begin{eqnarray}\label{eq:v3-2}
		a_{\al\be}=\frac{1}{c_3}(\frac{\kat\mu}{2}-2\rho k_3)\eta_{\al\be} &\Rightarrow&
		k_3=\frac{1}{2\rho}(\frac{\kat\mu}{2} -  \frac{c_3}{3}Tr(a^{\al\be}) )   \\
		c^{\al\be}=-2\frac{m_3\rho}{c_2}\eta^{\al\be}& \Rightarrow& m_3=-\frac{c_2}{6\rho}Tr(c^{\al\be})\label{eq:c-3-1}\\
		\ka\eta^{\al\be}+c^{\be\al}={-}\frac{\ka}{c_3\rho}\eta^{\al\be}&\Rightarrow& c^{\al\be}={-}\ka(\frac{1}{c_3\rho}+1)\eta^{\al\be}\label{eq:c-3-2}\\
		b_{\al\be\gamma}=0\label{4.26}\ .&&
		\end{eqnarray}
Compatibility between \eqref{eq:c-ab1}, (\ref{eq:c-3-1}) and (\ref{eq:c-3-2}) requires that
\begin{equation}\label{4.27}
		c^{\al\be}=\frac{\rho}{ c_1}\eta^{\al\be}=-2\frac{m_3\rho}{c_2}\eta^{\al\be}=-\ka(\frac{1}{c_3\rho}+1)\eta^{\al\be}\ .
\end{equation}
The holographic link between the 4D bulk theory $S_{tot}$ \eqref{eq:Stot} and the 3D boundary theory $S_{3D}$ \eqref{eq:S-gen} is realized if the coefficients appearing in this latter are		
		\begin{empheq}{align}
		k_3&=-\frac{c_3}{6\rho}\; Tr(a^{\al\be})+\frac{\mu\ka}{4\rho}\\
		m_3&=-\frac{c_2}{6\rho}\; Tr(c^{\al\be})\ .
		\end{empheq}
Therefore, the resulting 3D action, written in terms of parameters appearing in the 4D action $S_{tot}$ \eqref{eq:Stot}, reads
\begin{equation}\label{4.31}
		\Sa=\int d^3X\biggl[\frac{\kappa}{2\nu\rho}G_{\al\be}\Gt^{\al\be}+\bigl(\frac{\mu\ka}{4\rho}- 
		\frac{c_3}{6\rho}Tr(a^{\al\be})\bigr)\; \;\Gt_{\al\be}\Gt^{\al\be}
		-\frac{c_2}{6\rho}Tr(c^{\al\be})\; \epsilon^{\al\be\ga}\D_\al\lat_\be\lat_\ga\biggr]\ .
\end{equation}
The presence in the action $\Sa$ \eqref{4.31} of a topological Chern-Simons-like term, with a dimensional coefficient ($[c_2/\rho]=1$) and which can be switched off by requiring $c_2=0$, reminds us of the 3D Maxwell-Chern-Simons theory, where the coefficient of the Chern-Simons term serves as a topological mass for the gauge field. In Appendix A we compute the matrix formed by propagators of this theory, which involves two gauge fields, and we show that a similar mechanism of generation of a topological mass is not reproduced in this case.
The e.o.m of the action (\ref{4.31}) are
		\begin{empheq}{align}
		&\frac{\kappa}{\nu\rho}\; \D_\al\Gt^{\al\gamma}=0\label{eq:eom-la'}\\
		&\frac{\kappa}{\nu\rho}\;\D_\al G^{\al\gamma}+\biggl[\frac{\mu\ka}{\rho}-2\frac{c_3}{3\rho}Tr(a^{\al\be})\biggr]\;\D_\al\Gt^{\al\gamma}-\frac{c_2}{3\rho}\; Tr(c^{\al\be})\; \epsilon^{\al\be\gamma}\D_\al\lat_\be=0\ , \label{eq:eom-lat'}
		\end{empheq}
and, using (\ref{eq:eom-la'}), the above EOM become
		\begin{empheq}{align}
		&\D_\al\Gt^{\al\gamma}=0 \label{4.34}\\
		&\D_\al G^{\al\gamma}+\tilde m\; \epsilon^{\al\be\gamma}\D_\al\lat_\be=0  \label{4.35}\ ,
		\end{empheq}
where
\begin{equation}
\tilde{m}\equiv -\frac{c_2\nu}{3\kappa}Tr(c^{\al\be})\ .
\end{equation}
Some of the parameters 
 will be set by the request that the action \eqref{4.31} yields a positive definite energy density $T_{00}$, where $T_{\al\be}$ is the energy-momentum tensor of the theory.  This will be done in the next Section.  Finally, we remark that the same EOM \eqref{4.34} and \eqref{4.35} can be obtained from the action
\begin{equation}\label{Shat}
		\bar{S}^{(1)}_{3D}=\int d^3 X\ \biggl(\ka G_{\al\be}\Gt^{\al\be}+m\epsilon^{\al\be\gamma}\D_\al\lat_\be\lat_\gamma\biggr)\ ,
\end{equation}
with $m/\ka=\tilde{m}$. One might therefore wonder if the two 3D actions \eqref{4.31} and \eqref{Shat} are equivalent. This question belongs to the more general issue of the meaning of  equivalent field theories. The answer is that two theories can be considered equivalent if their physical observables coincide. Given that the physical observables in field theory are the Green functions, in Appendix A we show that the simplest Green functions derived from the actions \eqref{4.31} and \eqref{Shat}, $i.e.$ the two-point functions, a.k.a. the propagators, differ. Hence, we have here a nice example of two theories with equivalent EOM, but which are nonetheless physically inequivalent. 

\subsection{Case 2}\label{sec:c2}

The first linking equation \eqref{eq:cont4.1} of Case {\bf 2} concerns the EOM (\ref{eq:eom-xi}) and the BC (\ref{eq:bdc1'}), which we write here again
\begin{empheq}{align}
		&\epsilon_{\la\ka\al}\D^\la\bigl[\eta^{\al\be}\epsilon_{\gamma\delta\be}\D^\gamma(\rho\xi^\delta+\sigma\xit^\delta)\bigr]=0\\
		&\epsilon_{\be\gamma\delta}\D^\gamma [(\ka\eta^{\al\be}+c^{\be\al})\tilde{\xi}^\delta+2(a^{\al\be}+b^{\kappa\be\al}\D_\kappa)\xi^\delta]=0\ .
		\end{empheq}
We observe that \eqref{eq:cont4.1}  is satisfied if 
		\begin{empheq}{align}
		\ka\eta^{\al\be}+c^{\be\al}&=\frac{\sigma}{c_4}\eta^{\al\be}\qquad\Rightarrow\qquad c^{\al\be}=(\frac{\sigma}{c_4}-\ka)\eta^{\al\be}\label{eq:v2-1}\\
		a^{\al\be}&=\frac{\rho}{2c_4}\eta^{\al\be}\label{eq:c-4-4} \\
		b^{\be\al\gamma}&=0 \label{4.42}\ .
		\end{empheq}
The second linking equation involves the EOM \eqref{eq:eom-xi-2}
		\begin{equation}
		-2m_3\epsilon_{\la\al\theta}\D^\la(\rho\xi^\al+\sigma\xit^\al)+\epsilon_{\la\al\theta}\D^\la\bigl\{\epsilon_{\be\ga\delta}\D^\gamma\bigl[(\kat\nu-4k_3\sigma)\eta^{\al\be}\xit^\delta+(\kat\mu-4k_3\rho)\eta^{\al\be}\xi^\delta\bigr]\bigr\}=0
		\end{equation}
and the following combination of BC
		\begin{equation}
		c_5\;(\ref{eq:bdc1'})+c_6\;\mathrm{curl}(\ref{eq:bdc2'})\ ,
		\end{equation}
which explicitly reads
		\begin{equation}\label{eq:bdc-4}
		c_5\epsilon_{\be\gamma\delta}\D^\gamma\bigl[(\ka\eta^{\ka\be}+c^{\be\ka})\xit^\delta+2a^{\ka\be}\xi^\delta\bigr]+c_6c_{\al\be}\epsilon^{\ka\la\al}\D_\la(\epsilon^{\be\gamma\delta}\D_\gamma\xi_\delta)=0\ .
\end{equation}
The contact between 4D and 3D theories is achieved if 
\begin{eqnarray}\label{eq:v3-2}
		\kat\nu-4k_3\sigma=0\quad&\Rightarrow&\quad k_3=\frac{\kat\nu}{4\sigma}\ ,\quad\sigma\neq0\\
		a_{\al\be}=-\frac{m_3\rho}{c_5}\eta_{\al\be}
		\quad&\Rightarrow&\quad m_3=-\frac{c_5}{3\rho}Tr(a^{\al\be})
		\label{eq:c-4-5} \\
		c^{\al\be}=\frac{1}{c_6}(\kat\mu-4k_3\rho)\eta^{\al\be}&=\dfrac{\ka}{\sigma c_6}\eta^{\al\be}&\label{eq:c-4-1}\\
		\ka\eta^{\al\be}+c^{\be\al}=-\frac{2m_3\sigma}{c_5}\eta^{\al\be}\quad&\Rightarrow&\quad m_3=-\frac{\rho c_5}{2}(\ka+\frac{1}{3}Tr(c^{\al\be}))\label{eq:c-4-2}\ .
\end{eqnarray}
Compatibility between (\ref{eq:v2-1}), (\ref{eq:c-4-1}) and (\ref{eq:c-4-2}) requires that
		\begin{equation}
		\frac{\sigma}{c_4}-\ka=\frac{\ka}{\sigma c_6}=-\frac{2m_3\sigma}{c_5}-\ka
		\end{equation}
which translates in
		\begin{empheq}{align}
		c_5=\; &-2m_3 c_4 \label{eq:c-4-3}\\
		c_4=\; &\frac{\sigma^2c_6}{\ka(1+\sigma c_6)}\ ,
		\end{empheq}
and the relation \eqref{eq:c-4-3} is also confirmed by requiring compatibility between \eqref{eq:c-4-4} and \eqref{eq:c-4-5}. Notice that from \eqref{eq:c-4-5} and \eqref{eq:c-4-2} we get
		\begin{equation}
		Tr(a^{\al\be})=\frac{3}{2}\rho^2\bigl[\ka+\frac{1}{3}Tr(c^{\al\be})\bigr]
		\end{equation}
which is a constraint between parameters of the 4D theory, coming from the bulk-boundary correspondence, which is an interesting result.\\
Finally, the 3D action \eqref{eq:S-gen} which realizes the holographic contact through Case {\bf 2}, written entirely in terms of parameters of the 4D action $S_{tot}$ \eqref{eq:Stot}, is 
\begin{equation}\label{eq:S1gen}
		\Sb=\int d^3X\biggl(-\frac{\kappa}{2(\mu\sigma-\nu\rho)}G_{\al\be}\Gt^{\al\be}+ 
		\frac{\kappa}{(\mu\sigma-\nu\rho)}\frac{\nu}{4\sigma}\;\Gt_{\al\be}\Gt^{\al\be}-\frac{c_5}{3\rho}Tr(a^{\al\be})\epsilon^{\al\be\ga}\D_\al\lat_\be\lat_\ga\biggr)\ ,
\end{equation}
which is of the same type of $\Sa$ \eqref{4.31}, with a different choice of the parameters and with the same possibility of switching off the Chern-Simons term by putting $c_5=0$, thus proving the not obvious fact that the two apparently inequivalent Cases {\bf 1} and {\bf 2} yield indeed the same 3D theory, which therefore turns out to be uniquely determined by the holographic contact.
	
\section{Energy-momentum tensor}

On the boundary of 4D Maxwell action we found the following unique model, holographically compatible with the 4D bulk theory in a form and manner described in Section 4:
\begin{equation}
		S_{3D}=\int d^3X\ \Lag_{(b)}=\int d^3X\biggl(\kappa_1G_{\al\be}\Gt^{\al\be}+\kappa_2\Gt_{\al\be}\Gt^{\al\be}+m\epsilon^{\al\be\gamma}\D_\al\lat_\be\lat_\gamma\biggr)\ .
\label{5.1}\end{equation}
The coefficients $\ka_1$, $\ka_2$ and $m$ in \eqref{5.1} are expressed in terms of the parameters appearing in the bulk theory $S_{tot}$ \eqref{eq:Stot} according to the two possible ways of realizing the holographic contact described in Sections 4.1 and 4.2. The results are given by the actions $S^{(1)}_{3D}$ \eqref{4.31} and $S^{(2)}_{3D}$ \eqref{eq:S1gen}, which are both of the type \eqref{5.1}. We remark that it is not possible to reabsorb the massive coefficient of the Chern-Simons term in \eqref{5.1} by means of a rescaling of the fields $\lambda_\al(X)$ and $\tilde\lambda_\al(X)$. Hence, the dimensional parameter coupled to the Chern-Simons term in the action $S_{3D}$ \eqref{5.1} is the only true parameter of the theory.
A further necessary constraint on the parameters appearing in the action \eqref{5.1} comes from the energy density, $i.e.$ the  00-component of the energy-momentum tensor, which must be positive. 
The energy-momentum tensor is defined as
\begin{equation}\label{eq:Tmunu}
		T_{\al\be}=\frac{-2}{\sqrt{-g}}\frac{\delta S}{\delta g^{\al\be}}\ ,
\end{equation}
where we made explicit the dependence on the metric $g_{\al\be}$, which will be eventually put equal to the Minkowskian $\eta_{\al\be}$. Using the definition \eqref{eq:Tmunu}, the Chern-Simons term in \eqref{5.1} does not contribute, and we can forget about it in what follows. Writing the non-topological part of the action \eqref{5.1} as
\begin{equation}
		S_{nt}=\int d^3X\sqrt{-g}\ \bigl(\ka_1G_{\al\be}+\ka_2\Gt_{\al\be}\bigr)\Gt_{\ga\delta}\ g^{\al\ga}g^{\be\delta},
\end{equation}
we can apply the definition \eqref{eq:Tmunu} and, remembering that
\begin{equation}
\frac{\delta \sqrt{-g}}{\delta g^{\al\be}}=-\frac{1}{2}\sqrt{-g}\; g_{\al\be}\ ,
\end{equation}
we find
\begin{equation}
T_{\al\be}=
-2\ka_1(G_{\al\gamma}\Gt_\be^{\ \ga} + G_{\be\gamma}\Gt_\al^{\ \ga})
-4\ka_2\Gt_{\al\ga}\Gt_\be^{\ \ga}
+g_{\al\be}\bigl(\ka_1G_{\ga\delta}+\ka_2\Gt_{\ga\delta}\bigr)\Gt^{\ga\delta}\ .
\label{Tmunu}\end{equation}
As a check, we may calculate the trace of this energy-momentum tensor
\begin{equation}
T=g^{\al\be}T_{\al\be}=(D-4)(\ka_1G_{\al\be}\Gt^{\al\be}+\ka_2\Gt_{\al\be}\Gt^{\al\be})\ ,
\end{equation}
which vanishes for $D=4$, as it should. In Minkowskian spacetime $g_{\al\be}=\eta_{\al\be}$, the 00-component of \eqref{Tmunu} is
\begin{equation}
T_{00}=-2\ka_1G_{0i}\Gt_0^{\ i} -2\ka_2\Gt_{0i}\Gt_0^{\ i} 
-\ka_1G_{ij}\Gt^{ij}-\ka_2\Gt_{ij}\Gt^{ij}\ .
\label{T00}\end{equation}
As in Maxwell theory, we have to look for terms containing time derivatives of the fields, which must appear in the action with the positive sign, since, otherwise, sufficiently rapid change of the fields with time could always make the action $S_{3D}$ \eqref{5.1} a negative quantity with arbitrary large absolute value, and hence it could not have a minimum, as required by the principle of least action \cite{Landau}. The terms in \eqref{T00} containing time derivatives are
\begin{eqnarray}
T^{time}_{00}
&=&
-2\ka_1G_{0i}\Gt_0^{\ i} -2\ka_2\Gt_{0i}\Gt_0^{\ i} 
\\
&\simeq&
-2\ka_1(\partial_0\la_i\partial_0\lat^i -\partial_0\la_i\partial^i\lat_0 -  \partial_i\la_0\partial_0\lat^i) 
-2\ka_2(\partial_0\lat_i\partial_0\lat^i -\partial_0\lat_i\partial^i\lat_0 -  \partial_i\lat_0\partial_0\lat^i)\ .\nonumber
\end{eqnarray}
The terms with two time derivatives dominate, for fields rapidly varying with time. Hence it must be
\begin{equation}
\ka_1 < 0 \ ;\ \ka_2 < 0\ .
\end{equation}
Let us see what this implies for the Cases {\bf 1} and {\bf 2}  studied in Sections 4.1 and 4.2: 
\begin{description}
\item{Case {\bf 1}}\\

From \eqref{4.31} we have
\begin{empheq}{align}
\ka_1&=\frac{\ka}{2\nu\rho} <0 \quad\Rightarrow\quad   \nu\rho <0  \label{5.10}\\
\ka_2&=-\frac{c_3}{6\rho}\; Tr(a^{\al\be})+\frac{\mu\ka}{4\rho} <0\ . \label{5.11}
\end{empheq}
\item{Case {\bf 2}}\\

From \eqref{eq:S1gen} we have
\begin{eqnarray}
\ka_1=-\frac{\ka}{2(\mu\sigma-\nu\rho)} <0 &\Rightarrow&   \mu\sigma-\nu\rho >0  \label{5.12}\\
\ka_2= \frac{\ka}{(\mu\sigma-\nu\rho)}\frac{\nu}{4\sigma}<0 &\Rightarrow& \frac{\nu}{\sigma} <0\ , \label{5.13}
\end{eqnarray}
\end{description}
where  $\ka>0$ has been taken into account. 
Notice that in both cases the constraint \eqref{eq:condiz2} is automatically respected. This is very interesting, because it suggests that the unitarity of the Conformal Field Theories which are found on the boundary of 4D Maxwell theory is tightly related to the positivity of the energy density of the 3D theory found by means of the holographic contact discussed in Section 4. The conditions \eqref{5.10}, \eqref{5.11}, \eqref{5.12} and \eqref{5.13} have many solutions. In particular, solutions can be found which yield the same 3D action  $S_{3D}$ \eqref{5.1} and boundary action $S_{bd}$ \eqref{eq:Sbd}.\\ For instance in Case {\bf1} we can choose
\begin{equation}
\mu=-2\rho\ ;\ \nu=-\frac{1}{\rho}\ 
\label{5.14}\end{equation}
while in Case {\bf2}
\begin{equation}
\mu=0\ ;\ \nu=-\frac{1}{\rho}\ ;\ \sigma=\frac{1}{\rho}\ ,
\label{5.15}\end{equation}
these, together with the above request of matching solutions ($i.e.$ requiring same $a^{\al\be},\ b^{\al\be\ga},\ c^{\al\be}$ and $\ka_1,\ \ka_2,\ m$ for both cases), lead to constraints between the coefficients $c_i$ appearing in the linking equations \eqref{eq:cont3.1}, \eqref{eq:cont3.2}, \eqref{eq:cont4.1} and \eqref{eq:cont4.2}, which are:
\begin{equation}
c_1=\frac{c_6}{\ka}\ ;\ c_2=\rho (c_6+\rho)c_5\ ;\ c_3=-\frac{1}{\rho}\frac{c_6}{c_6+\rho}\ ;\ c_4=\frac{1}{\rho\ka}\frac{c_6}{c_6+\rho}\ .\label{5.16}
\end{equation}
Two parameters are left to choose: $c_6$ and $c_5$ (notice that the latter, taken equal to 0, allow us to switch off the CS term in the action). With the choice $c_6=2\rho$ and $c_5=\rho$, both cases correspond to the same 3D action \eqref{5.1} with
\begin{equation}
\ka_1=-\frac{\ka}{2}\ ;\ \ka_2=-\frac{\ka}{4}\ ;\ m=-\frac{3}{4}\ka\rho^2<0
\label{5.17}
\end{equation}
and the same boundary action \eqref{eq:Sbd} with
\begin{equation}
a^{\al\be}=\frac{3}{4}\ka\rho^2\eta^{\al\be}\ ;\ b^{\al\be\ga}=0\ ;\ c^{\al\be}=\frac{\ka}{2}\eta^{\al\be}\ .
\label{5.18}
\end{equation}

\section{Conclusions}

Undoubtedly, the most known and also physically relevant role played by boundaries concerns TFT, in particular in 3D and 4D. This fact constitutes a kind of interesting paradox: 
TFT, indeed,  are characterized by global observables of geometrical type only, vanishing Hamiltonian, no energy-momentum tensor and lack of particle interpretation. Nonetheless, when boundaries are introduced, TFT show  a surprisingly rich physical content, revealing themselves as the most promising low-energy effective field theories for phenomena, like the Fractional Quantum Hall Effect and the physics of the  Topological Insulators, which are not completely understood yet. The combination of non-physical topological bulk and rich physical boundary dynamics finds some deviation in 3D, where non-topological bulk terms have also been considered. On a completely different side, an important example of non-TFT with boundary is given by the gauge/gravity duality, where gravity with an AdS black hole metric in 5D has a Conformal Field Theory as 4D holographic counterpart. Quite unexpectedly, despite its original stringy framework and much later after its first appearance in Literature, the AdS/CFT correspondence found relevant physical applications in Condensed Matter Theory (again!), and, in particular, promising developments concern the theory of superconductivity and of strange metals. Driven also by this important example, we focused our attention on the introduction of a boundary in a purely non-TFT in 4D where, to our knowledge, it has not been studied yet if and which role is played by a boundary. This question motivated our paper, where the 4D Maxwell theory of electromagnetism, $i.e.$ a theory which does not need a boundary to display physical properties, has been considered in a half-space, with single-sided boundary. We summarize our results as follows
\begin{itemize}
\item 
The first point which should be stressed is that 4D Maxwell theory shows a non trivial boundary dynamics, which therefore is not peculiar to TFT, contrary to what usually is believed. There are however similarities and differences with respect to TFT.
\item 
On the boundary of 4D Maxwell theory the broken Ward identities \eqref{eq:wi1} and \eqref{eq:wi2} are found, which identify two conserved currents \eqref{subeq:cc1} and \eqref{subeq:cc2}. This reminds the physics of the surface states of the Topological Insulators in 3D, which suggests that an aspect to be developed in the future is to investigate whether the 4D Maxwell theory  might be seen as an effective bulk theory of the 3D Topological Insulators, alternative to the 4D topological BF models \cite{Cho:2010rk}. 
\item
By means of  \eqref{defB} it is possible to define the 3D field $B_\al(X)$  whose components form the Ka\c{c}-Moody algebra \eqref{eq:B0,Bi} with a central charge  proportional to the inverse of the Maxwell coupling. 
The parameters appearing in \eqref{defB} correspond to different central charges, as represented by \eqref{eq:condiz1}, each identifying a different Conformal Field Theory.
This is an important difference with respect to TFT, which are characterized by a one-to-one correspondence between bulk coupling constants and central charges.   
The relevant boundary algebra appears to be formed by the subset \eqref{defB} of the total number of components of the bulk fields. An identical mechanism  occurs in the topological twist of N=2 Super Yang-Mills Theories \cite{Witten:1988ze}. This is a curious analogy which deserves further deepening.
\item
We found that the 3D theory depends on two vector fields, it is gauge invariant and it must satisfy the relation \eqref{defp}, coming from the compatibility with the Ka\c{c}-Moody algebra \eqref{eq:B0,Bi}. These constraints exclude the possibility of having on the boundary of 4D Maxwell theory a purely TFT.
\item 
The holographic contact with the bulk theory is realized, as in TFT, by matching the equations of motion of the 3D boundary theory with the boundary conditions found for the bulk theory. The difference with the TFT case is that this contact can be realized in two non equivalent (and more complicated) ways. The non trivial result is that, no matter how the holographic contact is obtained, we land on the unique action \eqref{5.1}, which has not been studied previously. 
\item
The boundary term \eqref{eq:Sbd} is physically relevant and necessary, for at least two reasons. The first is that it determines the boundary conditions  \eqref{eq:bdc1} and \eqref{eq:bdc2}, which would be trivial without the boundary term. The second is that the couplings of the 3D action we find as ``holographic counterpart'' \eqref{4.31} (or \eqref{eq:S1gen} ) depend on the coefficients of the boundary term \eqref{eq:Sbd}. The 3D actions we find are non-trivial: they have non vanishing energy momentum tensor and Hamiltonian, which also depend on the boundary term, thus giving to it a physical meaning. 
\item
The action \eqref{5.1} describes two coupled photon-like vector fields, with a topological Chern-Simons term for one of them. We computed the propagators of the theory which show that, despite the similarity with the 3D Maxwell-Chern-Simons theory, a mechanism of topological mass generation does not take place in this case.
\item
The energy-momentum tensor \eqref{Tmunu} of the theory \eqref{5.1} reveals a non trivial physical content. In particular, we tuned the coefficients appearing in the 3D action in order to have a positive definite energy density. 
\item
The holographic dictionary \cite{Zaanen:2015oix} might be improved by an additional entry involving the unitarity of the Conformal Field Theory found on the boundary of 4D Maxwell theory and the positivity of the energy density of its 3D holographic counterpart, represented by the action \eqref{5.1}. In fact, asking that the 00-component of the energy-momentum tensor \eqref{T00} derived from the action \eqref{5.1} is positive, automatically implies that the central charge of the Ka\c{c}-Moody algebra \eqref{eq:B0,Bi} is positive as well, thus ensuring the unitarity of the corresponding Conformal Field Theory.
\end{itemize}

\section*{Acknowledgments}
We gratefully acknowledge Dario Ferraro and Paolo Solinas for a careful reading of the draft and for their enlightening observations.
\appendix

\section{Propagators} 

In Section 3 we have seen that the 4D Maxwell theory with planar boundary $x_3=0$ induces on its 3D boundary the action $S_{3D}$ \eqref{eq:S-gen}, which we report here:
\begin{equation}
		S_{3D}\equiv\int d^3X\ \Lag_{(b)}=\int d^3X\biggl(\kappa_1G_{\al\be}\Gt^{\al\be}+\kappa_2\Gt_{\al\be}\Gt^{\al\be}+m\epsilon^{\al\be\gamma}\D_\al\lat_\be\lat_\gamma\biggr)\ ,
\label{A.1}\end{equation}
and in Section 4 we showed that its EOM are compatible with the BC \eqref{eq:bdc1} and \eqref{eq:bdc2} of the bulk theory, for certain values of the coefficients $\kappa_1$, $\kappa_2$ and $m$. We see that in \eqref{A.1} a Chern-Simons-like topological term is present, coupled to $m_3$, which is a dimensionful parameter. The same term, when coupled to Maxwell theory, give rise to a topological mass \cite{Deser:1982vy}. In this Appendix we would like to explore whether a similar mechanism occurs for the action $S_{3D}$ \eqref{A.1} we found as the holographic counterpart of 4D Maxwell theory. In order to do that, we have to compute the propagators of this 3D theory, being mainly interested in its (possibly massive) poles. The necessary, preliminary step is to add to \eqref{A.1} a gauge fixing term
\begin{equation}
S^{(gf)}_{3D}=\int d^3 X\ \biggl(-\frac{1}{2\xi}(\D_\al\la^\al)^2-\frac{1}{2\xit}(\D_\al\lat^\al)^2\biggr)\ ,
\label{A.2}\end{equation}
where $\xi$ and $\tilde\xi$ are gauge parameters. In momentum space ($\D_\al\to-ip_\al$), the gauge fixed action
\begin{equation}
S_{3D}^{(tot)}[\la,\lat]=S_{3D}[\la,\lat]+S^{(gf)}_{3D}[\la,\lat]
\label{A.3}\end{equation}
reads
\begin{equation}
S_{3D}^{(tot)}[\hat\la,\hat\lat]=
\int d^3p\ \hat\la_A(p)K^{AB}(p)\hat\la_B(-p)\ ,
\label{A.4}\end{equation}
where we adopted the compact notation $\hat\la_A(p)\equiv(\hat\la_\al(p),\hat\lat_\alt(p))$ for the Fourier transforms of  the fields ($\la_\al(X),\ \lat_\al(X))$, and the indices $A\equiv(\al,\alt)$, $B\equiv(\be,\bet)$, $C\equiv(\ga,\gat)$, where the indices with the $\sim$ refers to the matrix element acting on $\lat$.
In \eqref{A.4}, the matrix $K^{AB}(p)$ is given by
\begin{equation}
		K^{AB}(p)\equiv\left(
			\begin{array}{cc}
			-\frac{1}{2\xi}p^\al p^\be & \ka_1(p^2\eta^{\alt\be}-p^\alt p^\be) \\
			 \ka_1(p^2\eta^{\alt\be}-p^\alt p^\be) & 2\ka_2p^2\eta^{\alt\bet}-(2\ka_2+\frac{1}{2\xit}) p^\alt p^\bet+im\epsilon^{\alt\bet\delta}p_\delta\\
			\end{array}
		\right).
		\end{equation}
The matrix $\Delta_{BC}(p)$ formed by the propagators, in its general form is
\begin{equation}
		\Delta_{BC}(p)=\left(
			\begin{array}{cc}
			\Delta^{_{(1)}}_ {\be\ga}(p)&\Delta^{_{(2)}}_{\bet\ga}(p)\\
			\Delta^{_{(2)}}_{\be\gat}(p)&\Delta^{_{(3)}}_{\bet\gat}(p)\\
			\end{array}
		\right)\ ,
		\end{equation}
where 
\begin{equation}
\Delta^{(i)}_{\al\be}(p)=A_i(p)\eta_{\al\be}+B_i(p)p_\al p_\be+iC_i(p)\epsilon_{\al\be\gamma}p^\gamma\ .
\label{A.7}\end{equation}
In \eqref{A.7}, $A_i(p)$, $B_i(p)$ and $C_i(p)$ are functions of $p^2$, and are determined by imposing that 
$\Delta_{BC}(p)$ is the inverse of $K^{AB}(p)$, $i.e.$~:
\begin{equation}\label{A.8}
		K^{AB}\Delta_{BC}=\delta^A_{\ C}=\left(
			\begin{array}{cc}
			\delta^\al_{\ \ga}&0\\
			0&\delta^\alt_{\ \gat}\\
			\end{array}
		\right)\ .
		\end{equation}
The matrix equation \eqref{A.8} can be easily solved to finally find the propagators of the theory described by $S_{3D}^{tot}[\la,\lat]$ \eqref{A.3}~:
\begin{empheq}{align}
&\Delta^{(1)}_{\al\be}(p)=\langle\la_\al\la_\be\rangle(p)=
-\frac{1}{\ka_1^2p^2}\biggl[2\ka_2\eta_{\al\be}+(2\xi\ka_1^2+2\ka_2)\frac{p_\al p_\be}{p^2}+im\frac{\epsilon_{\al\be\gamma}p^\gamma}{p^2}\biggr]
\label{A.9}\\
&\Delta^{(2)}_{\al\tilde\be}(p)=\langle\la_\al\lat_\be\rangle(p)=\frac{1}{\ka_1 p^2}\biggl(\eta_{\al\be}-\frac{p_\al p_\be}{p^2}\biggr)
\label{A.10}\\
&\Delta^{(3)}_{\tilde\al\tilde\be}(p)=\langle\lat_\al\lat_\be\rangle(p)=-2\xit\frac{p_\al p_\be}{(p^2)^2}\label{A.11}\ ,
		\end{empheq}
which do not show any massive pole, so that we can conclude that the presence of a topological term in the action $S_{3D}[\la,\lat]$ \eqref{A.1} does not induce any mechanism of generation of a topological mass like it happens in Maxwell-Chern-Simons theory in three spacetime dimensions.

As we remarked in Section 4.1, the EOM derived from the action $S_{3D}$ \eqref{A.1} are equivalent to those obtained from the action $\bar{S}^{(1)}_{3D}$ \eqref{Shat}, which we write here again
\begin{equation}\label{A.12}
		\bar{S}^{(1)}_{3D}=\int d^3 X\ \biggl(\ka G_{\al\be}\Gt^{\al\be}+m\epsilon^{\al\be\gamma}\D_\al\lat_\be\lat_\gamma\biggr)\ .
\end{equation}
We now compute the propagators for the theory described by this latter action, and we show that these, indeed,  do not coincide with those we computed for the action \eqref{A.3}, given by \eqref{A.9}, \eqref{A.10} and \eqref{A.11}. Hence, the two theories have at least one Green function (the simplest, $i.e.$ the two-point function) which differs. Therefore, we must conclude that the two theories do not have the same physical content, although their EOM are equivalent. After adding to the action \eqref{A.12} the same gauge fixing term \eqref{A.2}, the gauge fixed action 
\begin{equation}
\bar{S}_{3D}^{(tot)}[\la,\lat]=\bar{S}_{3D}[\la,\lat]+S^{(gf)}_{3D}[\la,\lat]
\label{A.13}\end{equation}
in Fourier transform is
\begin{equation}
\bar S_{3D}^{(tot)}[\hat\la,\hat\lat]=
\int d^3p\ \hat\la_A(p)\bar{K}^{AB}(p)\hat\la_B(-p)\ ,
\label{A.14}\end{equation}
where the matrix $\bar{K}^{AB}(p)$ is given by
\begin{equation}
		\bar{K}^{AB}(p)\equiv\left(
			\begin{array}{cc}
			-\frac{1}{2\xi}p^\al p^\be & \ka(p^2\eta^{\alt\be}-p^\alt p^\be) \\
			 \ka(p^2\eta^{\alt\be}-p^\alt p^\be) & im\epsilon^{\alt\bet\delta}p_\delta-\frac{1}{2\xit} p^\alt p^\bet\\
			\end{array}
		\right)\ .
\end{equation}
The propagator matrix $\bar{\Delta}_{BC}(p)$ must satisfy
\begin{equation}\label{A.16}
		\bar{K}^{AB}\bar{\Delta}_{BC}=\delta^A_{\ C}=\left(
			\begin{array}{cc}
			\delta^\al_{\ \ga}&0\\
			0&\delta^\alt_{\ \gat}\\
			\end{array}
		\right)\ ,
\end{equation}
and its most general form is 
		\begin{equation}
		\bar{\Delta}_{BC}(p)=\left(
			\begin{array}{cc}
			\bar{\Delta}^{_{(1)}}_ {\be\ga}(p)&\bar{\Delta}^{_{(2)}}_{\bet\ga}(p)\\
			\bar{\Delta}^{_{(2)}}_{\be\gat}(p)&\bar{\Delta}^{_{(3)}}_{\bet\gat}(p)\\
			\end{array}
		\right)\ ,
		\end{equation}
with
		\begin{equation}
		\bar\Delta^{(i)}_{\al\be}(p)=\bar{A}_i(p)\eta_{\al\be}+\bar{B}_i(p)p_\al p_\be+i\bar{C}_i(p)\epsilon_{\al\be\gamma}p^\gamma\ .
		\end{equation}
Analogously to what we already did, the matrix equation \eqref{A.16} is solved by the following propagators: 
\begin{empheq}{align}
&\bar\Delta^{(1)}_{\al\be}(p)=\langle\overline{\la_\al\la_\be}\rangle(p)=-\frac{1}{p^2}\biggl(2\xi\frac{p_\al p_\be}{p^2}+\frac{im}{\ka^2}\frac{\epsilon_{\al\be\gamma}p^\gamma}{p^2}\biggr)\label{A.19}\\
		&\bar\Delta^{(2)}_{\al\tilde\be}(p)=\langle\overline{\la_\al\lat_\be}\rangle(p)=\frac{1}{\ka p^2}\biggl(\eta_{\al\be}-\frac{p_\al p_\be}{p^2}\biggr)\\
		&\bar\Delta^{(3)}_{\tilde\al\tilde\be}(p)=\langle\overline{\lat_\al\lat_\be}\rangle(p)=-2\xit\frac{p_\al p_\be}{(p^2)^2}\ ,
		\end{empheq}
which, again, do not show any topologically generated massive pole and,  which matters more now, do not coincide with the propagators previously computed for the action $S^{tot}_{3D}$ \eqref{A.3}. In particular, the propagators $\Delta^{(1)}_{\al\be}(p)$ \eqref{A.9} and 
$\bar\Delta^{(1)}_{\al\be}(p)$ \eqref{A.19} differ. Hence, as anticipated, the two theories are physically inequivalent.

\section{Symmetries}

The presence of the boundary at $x_3=0$ does not prevent the bulk action \eqref{eq:Smax} from being invariant under the gauge transformation
\begin{equation}
\delta A_\mu=\partial_\mu\Phi\ ,
\end{equation}
where $\Phi(x)$ is a local gauge parameter. Under this respect, Maxwell theory differs form topological field theories like 3D Chern-Simons theory and BF models, whose Lagrangians transform into a total derivative. A common feature of all theories with boundary, however, is the partial breaking of general covariance, which justifies the axial gauge choice \eqref{gaugecond}, and of discrete Parity symmetry. Discrete symmetries are crucial for the boundary physics of topological field theories. Think for instance to the almost defining role of Time Reversal for the Fractional Quantum Hall Effect and for Topological Insulators in Chern-Simons and BF theories with boundary, respectively. In order to investigate whether a similar role is played in the unknown 3D theory, or theories, possibly induced on the boundary of 4D Maxwell theory, we pay a particular attention to discrete symmetries.  On the $x_\al$ coordinates, Parity $\Par$ and Time Reversal $\T$  are defined as follows
\begin{empheq}{align}
		&\Par x_0\to\quad x_0,\quad\Par x_i\to-x_i\\
		&\T x_0\to-x_0,\quad\T x_i\to\quad x_i\ .
		\end{empheq}
Correspondingly, on the fields $A_\al(X)$ and $\At_\al(X)$ we have
		\begin{empheq}{align}
		&\Par A_0\to\quad A_0,\quad\Par A_i\to-A_i\\
		&\T A_0\to-A_0,\quad\T A_i\to\quad A_i
		\end{empheq}

		\begin{empheq}{align}
		&\Par \At_0\to-\At_0,\quad\Par \At_i\to\At_i\\
		&\T \At_0\to-\At_0,\quad\T \At_i\to\At_i\ .
		\end{empheq}
In order for the boundary term $S_{bd}$ \eqref{eq:Sbd} to be $\Par$ and/or $\T$ invariant, we should impose the following constraints on the constant parameters appearing in \eqref{eq:Sbd}
		\begin{empheq}{align}
				\Par S_{bd}\to S_{bd}\quad&\Leftrightarrow \quad a^{0i}=b^{00i}=b^{ijk}=c^{00}=c^{ij}=0\label{eq:P}\\
				\T S_{bd}\to S_{bd}\quad&\Leftrightarrow\quad a^{0i}=b^{0ij}=b^{ij0}=c^{i0}=c^{0i}=0\label{eq:T}\\
				\Par\T S_{bd}\to S_{bd}\quad&\Leftrightarrow\quad b^{\al\be\gamma}=c^{\al\be}=0\ .\label{eq:PT}
		\end{empheq}
Now, for what concerns the parameters found in Section 5, which we report again here: 
\begin{eqnarray}
a^{\al\be} &=& \frac{3}{4}\ka\rho^2\eta^{\al\be} \label{B.11}\\
b^{\al\be\gamma} &=& 0 \label{B.12}\\
c^{\al\be} &=& \frac{\ka}{2}\eta^{\al\be}\ , \label{B.13}
\end{eqnarray}
it is readily seen that \eqref{B.11} is compatible with the request that $S_{bd}$ \eqref{eq:Sbd} satisfies both $\Par$ and $\T$ (\eqref{eq:P} and \eqref{eq:T}), while \eqref{B.13} is compatible only with \eqref{eq:T} $i.e.$ $\T$.

\end{document}